\documentclass[superscriptaddress,secnumarabic,amssymb,amsmath,nobibnotes,aps,prd,showkeys,showpacs,nofootinbib,preprint]{revtex4-1}

\usepackage{graphicx}
\usepackage{float}
\usepackage{bm}
\usepackage{amsmath}
\usepackage{amsfonts}
\usepackage{amssymb}
\usepackage{epstopdf}
\usepackage{natbib}%
\newcommand{\bee}{\begin{equation}}
\newcommand{\eee}{\end{equation}}
\newcommand{\eaa}{\end{eqnarray}}
\newcommand{\baa}{\begin{eqnarray}}
\def\ni{\noindent}

\usepackage{color}

\begin{document}

\title{\Large On the equipartition theorem and black holes \\ nongaussian entropies}

\author{Everton M. C. Abreu}\email{evertonabreu@ufrrj.br}
\affiliation{Departamento de F\'{i}sica, Universidade Federal Rural do Rio de Janeiro, 23890-971, Serop\'edica, RJ, Brazil}
\affiliation{Departamento de F\'{i}sica, Universidade Federal de Juiz de Fora, 36036-330, Juiz de Fora, MG, Brazil}
\affiliation{Programa de P\'os-Gradua\c{c}\~ao Interdisciplinar em F\'isica Aplicada, Instituto de F\'{i}sica, Universidade Federal do Rio de Janeiro, 21941-972, Rio de Janeiro, RJ, Brazil}

\author{Jorge Ananias Neto}\email{jorge@fisica.ufjf.br}
\affiliation{Departamento de F\'{i}sica, Universidade Federal de Juiz de Fora, 36036-330, Juiz de Fora, MG, Brazil}

\author{Ed\'esio M. Barboza Jr.}\email{edesiobarboza@uern.br}
\affiliation{Departamento de F\'isica, Universidade do Estado do Rio Grande do Norte, 59610-210, Mossor\'o, RN, Brazil}

\author{Albert C. R. Mendes}\email{albert@fisica.ufjf.br}
\affiliation{Departamento de F\'{i}sica, Universidade Federal de Juiz de Fora, 36036-330, Juiz de Fora, MG, Brazil}

\author{Br\'aulio B. Soares}\email{brauliosoares@uern.br}
\affiliation{Departamento de Ci\^encia e Tecnologia, Universidade do Estado do Rio Grande do Norte, Natal, RN, Brazil}



\begin{abstract}
\ni  In this Letter we have shown that, from the standard thermodynamic functions, the mathematical form of an equipartition theorem may be
related to the algebraic expression of a particular entropy initially chosen to describe the black hole event horizon. Namely, we have different equipartition 
expressions for distinct statistics. To this end, four different mathematical expressions for the entropy have been selected to demonstrate our objective.
Furthermore, a possible phase transition is observed in the heat capacity behavior of the  Tsallis and Cirto entropy model.

\end{abstract}

\maketitle

\section{introduction}

Black holes are among the most interesting and charming objects that constitutes the main targets of quantum gravity analysis.  The origin of a black hole, a collapsing star, for example, makes us to see that the geometry of short-distance fluctuations can be augmented to macroscopic measures.   This connection between high-energy and low-energy physics can bring serious questions about these systems dynamics at the Planck scale dimensions, which has consequences for the low-energy experiments.

The pioneering papers of Hawking and Bekenstein on black holes physics  \cite{swh,jdb} lead to a profound connection
between gravity and thermodynamics. We can mention, for example, that quantities as entropy and temperature can be associated with
the black hole horizon. Recently, some derivations of Bekenstein-Hawking entropy were constructed \cite{barrow,nosso-1,nosso-2,saridakis}.

The aim of this paper is to show that the algebraic expression of an equipartition theorem will depend firstly on the entropy initially chosen to describe the black hole event horizon. To this end, we will use thermodynamic functions, which are normally used in the black holes physics, that are the entropy and the temperature. Throughout the paper we use $\hbar=c=k_B=1$. In the context of the usual black hole area entropy law (BHAEL), $S=A/4G$,
it will be assumed that the number $N$ of degrees of freedom (DF) of the horizon satisfy the standard equipartition law\cite{pad}

\begin{eqnarray}
\label{equi}
M= \frac{1}{2} N T \,,
\end{eqnarray}

\ni where $T$ is the temperature and $M$ means the mass of black hole. To better organize our paper, we will follow a sequence such that  in the second section we have explained our procedure and consequently derived the usual equipartition theorem. In section three we have presented the Tsallis and Cirto black hole entropy model \cite{tc,mora} and the respective equipartition law. In section four the modified R\'enyi entropy \cite{bc,vh,mora2} was introduced and the corresponding equipartition theorem was derived. In section five the equipartition theorem in the black hole horizon for the Sharma-Mittal entropy\cite{sm1,sm2} is derived. The conclusions and final remarks were given in the last section.


\section{Black hole area entropy}

The thermodynamics of black holes are defined on the basis of the concepts of both the entropy
and temperature of black hole \cite{jdb,swh,wald}. The temperature of a black hole horizon is directly proportional to
its surface gravity. In Einstein gravitation theory, the horizon entropy of a black hole is proportional to its
horizon area, i.e., the entropy area law of a black hole.

Our first issue will be the Schwarzschild black hole entropy, which will describe the horizon. As we have mentioned in the introduction the BHAEL can be connected to the 
area $A$ of the horizon through the well known relation
\begin{eqnarray}
\label{sa}
S_{BH}= \frac{A}{4G} \,,
\end{eqnarray}

\ni where $G$ is the gravitational constant. In black hole physics, the area $A$ in Eq. (\ref{sa}) can be connected to the source mass $M$ by the relation
\begin{eqnarray}
\label{am}
A=16 \pi G^2 M^2 \, .
\end{eqnarray}

\ni Consequently the entropy, Eq. (\ref{sa}), can be written as 
\begin{eqnarray}
\label{sm-0}
S_{BH}= 4 \pi G M^2 \,.
\end{eqnarray}

\ni The temperature is given by
\begin{eqnarray}
\label{sm}
\frac{1}{T}=\frac{\partial S(M)}{\partial M} \,,
\end{eqnarray}

\ni and using Eq. \eqref{sm-0} we have that
\begin{eqnarray}
\label{ts2}
\frac{1}{T}=\frac{\partial S_{BH}(M)}{\partial M}=8 \pi G M \,.
\end{eqnarray}

\ni The number $N$ of DF in the horizon can be given by assuming the relation \cite{ko}
\begin{eqnarray}
\label{ns}
N=4S\,,
\end{eqnarray}

\ni where $S$ is an specific entropy describing the horizon. So, using Eq. (\ref{ns}) in our initial case, we have 
\begin{eqnarray}
\label{ns2}
N=16\pi G M^2 \,.
\end{eqnarray}

\ni Combining Eqs. (\ref{ts2}) and (\ref{ns2}) and making some algebra then we can derive the usual equipartition theorem 
\begin{eqnarray}	
\label{eqs1}
M=\frac{1}{2} N T \,,
\end{eqnarray}

\ni which corresponds to the horizon energy.

To establish the physical coherence a standard test is to calculate the heat capacity of the model.   The sign of the heat capacity can support us in determining the stability of black holes.  Namely, a positive heat capacity is meaningful.   On the other hand, a negative heat capacity in such system shows a thermodynamical unstableness.

The heat capacity can be computed from the expression
\bee
\label{heat-capacity}
C\,=\,-\,\frac{[S_{BH}^{\prime}(M)]^2}{S_{BH}^{''}(M)} \,\,,
\eee

\ni where the prime means a derivative relative to $M$.  So, substituting the entropy of Eq. \eqref{sm-0} into Eq. \eqref{heat-capacity} we have that 
\bee
\label{h-capacity}
C_{BH}\,=\,-\,8 \pi G M^2 \,\,.
\eee

\ni which means, as well known, that a black hole is thermally unstable.    The negative heat capacity in this regime means that a slight drop in black hole temperature will cause an additional drop as energy is absorbed. The mechanism will continue forever and the black hole will keep feeding on the surrounding heat bath. On the other hand, a slight rise in black hole temperature from the equilibrium value will cause the black hole to radiate some net energy.   In this way, it will increase its temperature still further.   It will, eventually, conduct to the explosive vanishing of the black hole altogether.


\section{Tsallis-Cirto entropy}

Our second case will be the Tsallis and Cirto entropy\cite{tc}. In few words, we can say that entropies that generalize the Boltzmann-Gibbs one are necessary to retrieve the thermodynamics extensivity of nonstandard scenarios.   These authors have proposed a modified form of BHAEL such that
\begin{eqnarray}
\label{sat}
S_\delta=\gamma \, A^\delta \,,
\end{eqnarray}

\ni where $\gamma$ is an unspecified constant and $\delta$ denotes the non-additivity parameter. We can see that in the limit $\delta=1$ and $\gamma=1/4G\,$ the BHAEL is recovered. The cosmological implications of the modified form of BHAEL, Eq. (\ref{sat}), can be found, for example, in reference  \cite{ko2}. Using Eq. (\ref{am}) we can write Eq. (\ref{sat}) as
\begin{eqnarray}
\label{sat2}
S_\delta=\gamma' \, M^{2\delta} \,\,,
\end{eqnarray}

\ni where $\gamma'=\gamma (16 \pi G^2)^\delta$. Making use of Eqs. (\ref{sm-0}) and (\ref{ns}) we can derive, respectively, the bound temperature $T$ and the number $N$ of DF in the horizon such that
\begin{eqnarray}
\label{tst}
T=\frac{1}{2\, \delta\, \gamma'\, M^{2\delta-1}}\,\,,
\end{eqnarray}

\ni and

\begin{eqnarray}
\label{nst}
N=4 \, \gamma' \, M^{2\delta} \,\,.
\end{eqnarray}

\ni So, combining Eqs. (\ref{tst}) and (\ref{nst}) and after some algebra then the equipartition theorem in Tsallis-Cirto entropy takes the form
\begin{eqnarray}
\label{eqst}
M=\frac{1}{2}\, \delta \, N T \,\,,
\end{eqnarray}

\ni which corresponds to the horizon energy in Tsallis and Cirto entropy model. From Eq. (\ref{eqst}) we can observe the appearance of an extra term $\delta$ in the usual equipartition theorem, Eq. (\ref{eqs1}). When we make $\delta=1$ we recover
the usual equipartition law.

The heat capacity, using Eqs. \eqref{sat2} and \eqref{heat-capacity} is
\bee
\label{heat-cap-tsallis-cirto}
C_\delta\,=\,\frac{2\delta\gamma (16\pi G^2)^{\delta}}{1-2\delta}\,M^{2\delta} \,.
\eee


\ni When we make $\delta=1$ and $\gamma=1/4G$ in Eq. (\ref{heat-cap-tsallis-cirto}) we recover the heat capacity, Eq.(\ref{h-capacity}). 
.



\section{Modified R\'enyi entropy}

The third example that will be analyzed is the modified R\'enyi entropy. This model was suggested  by Bir\'o and Czinner and they considered the BHAEL
as a nonextensive Tsallis entropy \cite{tsa1,tsa2}. Next they wrote the R\'enyi entropy as a function 
of Tsallis entropy. The final form is given by
\begin{eqnarray}
\label{sar}
S_R= \frac{1}{\lambda} \, \ln (1+\lambda S_{BH}) \,\,,
\end{eqnarray}

\ni where $\lambda$ is a constant parameter. If we take the limit $\lambda\rightarrow 0$ in Eq. (\ref{sar}) then we have 
$S_R=S_T$ where $S_T$ is the Tsallis entropy. Using Eq. (\ref{am}) into Eq. (\ref{sar}) we have
\begin{eqnarray}
\label{sarm}
S_R= \frac{1}{\lambda} \, \ln \Big( 1+4 \pi \lambda G M^2 \Big) \,\,.
\end{eqnarray}

\ni Using Eqs. (\ref{sm-0}) and (\ref{ns}) we obtain, respectively, the black hole temperature and the number $N$ of DF as
\begin{eqnarray}
\label{tsr}
T =\frac{1+4 \pi \lambda G M^2}{8 \pi G M} \,\,,
\end{eqnarray}


\ni and
\begin{eqnarray}
\label{nr}
N=\frac{4}{\lambda} \ln \Big(1+4 \pi \lambda G M^2 \Big) \,.
\end{eqnarray}

\ni Using Eqs. (\ref{tsr}) and (\ref{nr}), after some algebra, we have the equation
\begin{eqnarray}
\label{eqsr}
M^2\bigg(1+\frac{1}{e^{\frac{N\lambda}{4}} - 1}\bigg)-\frac{2T}{\lambda}\,M =0 \,\,,
\end{eqnarray}

\ni  which non-zero solution is the mathematical expression for the equipartition theorem compatible with R\'enyi entropy that is
\begin{eqnarray}
\label{eqsr2}
M=\frac{2 T}{\lambda}  \frac{e^{\frac{N\lambda}{4}} - 1}{e^{\frac{N\lambda}{4}}}\,\,.
\end{eqnarray}

\ni It is straightforward to show that if we take the limit $\lambda\rightarrow 0$ in  Eq. (\ref{eqsr2}) then we recover the usual equipartition law, 
$M= 1/2 N T$. 



The heat capacity, using Eqs. \eqref{sarm} and \eqref{heat-capacity} is 
\bee
\label{heat-cap-mod-renyi}
C_R \,=\,\frac{8\pi G M^2}{4\pi \lambda G M^2 -1} \,\,.
\eee

\ni When we make $\lambda=0$ in (\ref{heat-cap-mod-renyi}) we recover the heat capacity, Eq.(\ref{h-capacity}). Here it is important to mention that Eq.(\ref{heat-cap-mod-renyi}) has already been obtained in ref. \cite{vh}.



\section{The Sharma-Mittal entropy}

R\'enyi and Tsallis entropies are formulations where the probability distributions are substituted by power-law distributions and the results are the so-called generalized entropies. The Sharma-Mittal (SM) entropy is a combined generalization of both Tsallis and R\'enyi entropies. The SM entropy leads us to relevant results in the cosmological scenario such as the description of the current accelerated Universe by using conveniently the vacuum energy. For more reference see \cite{asj} and references therein.

Although it is common procedure to use NE entropies to analyze BH thermodynamics properties, it is very different to consider the SM entropy for it.   Hence, it is completely new to investigate its equipartition law the way we do here.


We are interested here in the implications of the SM entropy in the equipartition law, which  was not taken in consideration until now.  The SM entropy is written as
\bee
\label{sm}
S_{SM}\,=\,\frac 1R \Big[(1\,+\,\delta S_T )^{\frac R\delta}\,-\,1 \Big]\,,
\eee

\ni where $S_T = A/4G\,$. $S_T$ is the nonextensive Tsallis entropy \cite {tsa1,tsa2}, as we mentioned in the last section, and $R$ and $\delta$ are free parameters. At the limits, $R \rightarrow 0$ and $R \rightarrow \delta$, the R\'enyi and Tsallis entropies, respectively, can be recovered.   It is important to explain that, although the Bekenstein-Hawking entropy is not an extensive entropy, it can be considered as a proper candidate for  $S_T$ in gravitational and cosmological scenarios \cite{nosso}, which is exactly the case here.   This choice is in total agreement with ref. \cite{tc}.

Having said that, using Eq. (\ref{sm-0}) we obtain the black hole SM temperature
\begin{eqnarray}
\label{tsm}
T =\frac{1}{8 \pi G M \, \left( 1+ 4 \pi \delta G M^2 \right)^{\frac{R}{\delta}-1} } \,\,.
\end{eqnarray}

As we have proceed in the last sections, the modified equipartition theorem for the SM statistics is
\bee
\label{sm-equip-law}
M\,=\, \frac{2}{\delta} \left[ \left(1+\frac{R}{4} N\right)^{\frac{\delta}{R}}-1\right] 
\left[ \left(1+\frac{R}{4} N\right)^{\frac{\delta}{R}}\right]^{\frac{R}{\delta}-1} \, T\,.
\eee
which is the energy of the horizon in the generalized case, i.e., when $R \neq \delta$.  It is clear that when $R = \delta$ we obtain the standard $M=1/2\,NT$ equipartition law. 

The heat capacity, using Eqs. \eqref{sm} and \eqref{heat-capacity} is
\bee
\label{hcsm}
C_{SM}\,=- \, \frac{ 8\pi G M^2 \left( 1+4 \pi \delta G M^2\right)^{\frac{R}{\delta}}}        { 1+ 8 \pi R G M^2 - 4 \pi \delta G M^2}\,\,,
\eee

\ni   When we make  $R=\delta$ in (\ref{hcsm}) we recover the heat capacity, Eq.(\ref{h-capacity}). It is important to comment here that Eq.(\ref{hcsm}) has already been obtained in ref. \cite{gz}.



\section{conclusions}

In this letter we have shown that the algebraic form of the equipartition theorem may depend on the particular form of the entropy initially chosen to describe
the black hole event horizon. Four cases have been studied in this paper, which are the usual black hole entropy, the Tsallis-Cirto entropy, the modified R\'enyi entropy 
and the Sharma-Mittal entropy,  For each case, the temperature
of the horizon and the DF number are different and the result is that the equipartition theorems corresponding to each model will also be logically different. 
Hence, we should pay attention to the fact that the 
algebraic form of the equipartition theorem is very sensible to the entropy formula initially chosen to describe the horizon. Therefore, important results, which are Eqs. (\ref{eqst}), (\ref{eqsr2}) and (\ref{sm-equip-law}) describe the equipartition theorem in black hole event horizon for different entropies models. Finally, we would like to mention that it is possible to obtain a black hole thermally stable state in the Tsallis and Cirto entropy model for $ \delta < 1/2$ in Eq.(\ref{heat-cap-tsallis-cirto}).

\section*{Acknowledgments}

\ni The authors thank CNPq (Conselho Nacional de Desenvolvimento Cient\' ifico e Tecnol\'ogico), Brazilian scientific support federal agency, for partial financial support, Grants numbers  406894/2018-3 (E.M.C.A.) and 303140/2017-8 (J.A.N.).

\end{document}